# Electronic structure and magnetic properties of $Ni_{0.2}Cd_{0.3}Fe_{2.5-x}Al_xO_4$ (0≤x≤0.4) ferrite nanoparticles

Shalendra Kumar[1*], Khalid Mujasam Batoo[2], S. Gautam[3], B. H. Koo[1], Alimuddin[2], K. H Chae[3], Chan Gyu Lee[1]

[1]School of Nano and Advanced Materials Engineering, Changwon National University, 9 Sarim dong, Changwon-641-773, Republic of Korea

[2]Department of Applied Physics, Aligarh Muslim University, Aligarh, UP 202002, India

[3]Nano Materials Analysis Center, Korea Institute of Science and Technology, Seoul 136-791, Republic of Korea

Structural, magnetic and electronic structural properties of $Ni_{0.2}Cd_{0.3}Fe_{2.5-x}Al_xO_4$ ferrites nanoparticles have been studied x-ray diffraction (XRD), transmission electron microscopy (TEM), dc magnetization, and near edge x-ray absorption fine structure spectroscopy (NEXAFS) measurements. Nanoparticles of $Ni_{0.2}Cd_{0.3}Fe_{2.5-x}Al_xO_4$ (0≤x≤0.4) ferrite were synthesized using sol-gel method. The XRD and TEM measurements show that all samples have single phase nature with cubic structure and have nanocrystalline behavior. From the XRD and TEM analysis, it is observed that particle size increases with Al doping. DC magnetization measurements infer that magnetic moment decreases whereas blocking temperature increases with increase in Al doping. It is observed that the magnetic moment decreases with Al doping which may be due to the dilution of the sublattice by the doping of Al ions. The NEXAFS measurements performed at room temperature indicates that Fe exist in mixed valence state.

Keywords: Ferrite Nanoparticles, XRD, TEM, DC Magnetization, NEXAFS

Corresponding author

E-mail: shailuphy@gmail.com (S. Kumar); chglee@changwon.ac.kr (C. G. Lee)

Ph: +82-55-213-3703; Fax: +82-55-261-7017

## 1. INTRODUCTION

In the recent years, magnetic nanoparticles have been a subject of great interest from both technological and fundamental points of view.[1-7] Among the magnetic materials, ferrites nanoparticles found to play crucial roles in the fast-packed miniaturization of modern electronic devices and biomedical applications. The origin for the spacious applications is related to the diversity of transition metal cations which can be integrated into the lattice of the parent magnetic structure. Particularly, mixed spinel ferrite nanoparticles have generated a large research effort because their magnetic properties differ markedly from single ferrite. Spinel ferrites crystallize into a cubic close packed structure of oxygen ions. The cations occupy two types of interstitial sites. One of them is called a tetrahedral (A) site with the cation surrounded by the four oxygen ions in tetrahedral coordination. The other interstitial position is known as octahedral (B) site with cation coordinated by six oxygen ions in octahedral symmetry.[8] In general, the cation distribution in spinel lattice has the form: $(D_{1-x}T_x) [D_xT_{2-x}] O_4$, where D and T are divalent and trivalent ions, respectively and x is so called the degree of inversion. The round and square brackets denote the cations located at the center of the tetrahedral lattice of oxygen (A) and those at the octahedral (B) lattice, respectively. The main source of the magnetic properties of spinel ferrites is the spin magnetic moment of the unpaired 3d electrons of the transition metal cations coupled by the superexchange interaction via the oxygen ions separating them. The magnetic properties of spinel ferrites, such as transition temperature and saturation magnetic moment are strongly dependent on the distribution of cations, type of doping atom and crystallite size. Therefore, the magnetic properties of these materials can be tailored by using the doping of the different transition metal cations as well as by varying the particle size.

The X-ray absorption spectroscopy (XAS) is recognized to be a powerful and most insightful technique for probing the chemical valency of the cations in the transition metal compounds.[6] The sensitivity of this technique to a crystal electric field with specific point group symmetry can be used to extract a number of significant parameters, such as crystal field strength (10 Dq) and hybridization etc. from a single experiment. The NEXAFS has been widely used to identify the valence states of ions in a material. In this procedure, the photons of specific characteristic energy are absorbed to produce the transition of a core electron to an empty state above the Fermi level which is governed by a set of dipole selection rules. In the present work, we have study the effect of Al doping on structural, magnetic and electronic structural properties of $Ni_{0.2}Cd_{0.3}Fe_{2.5}O_4$ ferrite nanoparticles.

## 2. EXPERIMENTAL DETAILS

Sol-gel technique has been used to fabricate the nanoparticles of Al doped $Ni_{0.2}Cd_{0.3}Fe_{2.5}O_4$ ferrite. The analytical grade chemical reagents used to prepare these materials were $Ni(NO_3)_2 \cdot 6H_2O$, $Cd(NO_3)_2 \cdot 2H_2O$, $Al(NO_3)_3 \cdot 9H_2O$, and $Fe(NO_3)_2 \cdot 9H_2O$. Details of the preparation of these materials are reported elsewhere.[9] The stoichiometric amounts of metal nitrates were dissolved in deionized water and then few drops of ethyl alcohol were added to this solution. In order to get the fine crystalline particles, few drops of N,N-dimethylformamide $C_3H_7NO$ (M.W. 73.10) were also added in the solution. The solution was put on the magnetic stirrer at 75 °C with constant stirring until the gel was obtained. The gel formed was annealed at 90 °C for 19 h followed by grinding for half an hour. Finally the powder was calcinated at 400 °C for 36 h to remove any organic material present in the system and then grinded for half an hour.

The prepared nanoparticles were characterized using X-ray diffraction (XRD), transmission electron microscopy (TEM), dc magnetization and near edge x-ray absorption fine structure spectroscopy (NEXAFS) measurements. Philips x-pert x-ray diffractometer with Cu Kα (λ = 1.54 Å) was used to study single phase nature of the samples at room temperature. TEM measurements were performed using FE-TEM (JEM 2100F). DC magnetization measurements were performed using Quantum Design physical properties measurement setup. The NEXAFS measurements of these samples, along with reference compounds of FeO ($Fe^{2+}$), $Fe_2O_3$ ($Fe^{3+}$) and $Fe_3O_4$ (Fe2+/Fe3+), at Fe $L_{3,2}$-edges were performed at the soft X-ray beam line 7B1 XAS KIST (Korea Institute of Science and Technology) of the Pohang Accelerator Laboratory (PAL), operating at 2.5 GeV with a maximum storage current of 200 mA. The spectra were simultaneously collected in the total electron yield (TEY) mode and the fluorescence yield (FY) mode at room temperature in a vacuum of better than $1.5 \times 10^{-8}$ Torr. The spectra in the two modes turned out to be nearly identical, indicating that the systems were so stable that the surface contamination effects were negligible even in the TEY mode. The spectra were normalized to the incident photon flux, and the energy resolution was better than 0.2 eV. The data were normalized and analyzed using Athena 0.8.054.

3. **RESULTS AND DISCUSSIONS**

Fig. 1 represents the θ-2θ XRD patterns of the $Ni_{0.2}Cd_{0.3}Fe_{2.5-x}Al_xO_4$ (0≤x≤0.4) nanoparticles. The reflection peaks observed from the samples can be well indexed with the standard pattern of the cubic spinel ferrites. The broadening of the reflection peaks suggests that samples have nanocrystalline behavior. The particle size of $Ni_{0.2}Cd_{0.3}Fe_{2.5-x}Al_xO_4$ (0≤x≤0.4) nanoparticle was calculated from the most intense peak (311) of XRD data using Debye Scherrer formalism[10]

$$t = 0.9\lambda / \beta \cos\theta \qquad 1$$

where $\beta = (\beta_M^2 - \beta_i^2)^{1/2}$. Here $\lambda$ is x-ray wavelength (1.54 Å for Cu K$_\alpha$), $\beta_M$ and $\beta_i$ is the measured and instrumental broadening in radians respectively and $\theta$ is the Bragg`s angle in degrees. The calculated value of particle size was found to increase from 4 to 7 nm with Al doping, which implies that Al doping favor the particle growth.

The size distributions and presence of any impurity phase were further studied by TEM images and electron diffraction patterns. Image J 1.3.2 J software was used to determine the average particle size and the size distribution, by analyzing approximately 200 particles. Figure 2 (a) and (b) displays TEM micrographs of $Ni_{0.2}Cd_{0.3}Fe_{2.5-x}Al_xO_4$ nanoparticles for x = 0.0 and 0.4. It is clearly evident from the micrographs that the prepared samples are composed of nanoparticles. From the particle size distribution histograms (see upper right inset in Fig. 2 (a) and (b)), the average diameters are in the range from 3.5 – 7.5 nm for different Al concentrations. The particle size measured from the TEM micrographs is in excellent agreement with the calculated values by Scherrer's formula. Upper left inset in Fig. 2 (a) and (b) shows the selected area electron diffraction (SAED) patterns observed from the $Ni_{0.2}Cd_{0.3}Fe_{2.5-x}Al_xO_4$ (0≤x≤0.4) nanoparticles. Rings in these patterns indicate clearly the randomly oriented single crystals and hence ruled out the presence of any impurity phase. SAED pattern (see upper left inset in Fig. 2 (a) and (b)) demonstrate that each nanoparticle is indeed in single phase.

Fig. 3 presents the magnetic hysteresis loop measurements of Al doped $Ni_{0.2}Cd_{0.3}Fe_{2.5}O_4$ (0≤x≤0.4) ferrite nanoparticles at room temperature. An absence of hysteresis, almost immeasurable coercivity and remanence represents the characteristic of super-magnetic behavior of the samples. In order to get more insight of the magnetic behavior of Al doped $Ni_{0.2}Cd_{0.3}Fe_{2.5}O_4$ nanoparticles, we had performed zero field cooled (ZFC) and field cooled (FC) magnetization. Inset in Fig. 3 shows the ZFC and FC magnetization measurements curve for x =

0.0 and 0.4. In the ZFC cycle, the sample was cooled from 320 to 20 K in the absence of magnetic field and after stabilization of the temperature a magnetic field of 500 Oe was applied. The data were then recorded while heating the sample. In FC cycle the sample was cooled from 320 to 20K in the presence of a magnetic field of 500 Oe and then the measurements were carried out while heating the sample in the same field. The presence of the cup in the ZFC and the bifurcation between ZFC and FC curve at certain temperature shows the characteristic feature of the superaparamagnetic system. The presence of bifurcation in the ZFC and FC curve at certain temperature shows the characteristic feature of a superparamagnetic behavior.[11,12] The temperature at which magnetization start decreasing is known as blocking temperature ($T_B$). The broad maxima observed in ZFC curve (denoted as $T_B$) indicates a certain particle size distribution in the system. Such type of magnetic behavior can be explained in the light of Stoner-Wohlfarth theory. According to Stoner-Wohlfarth theory, the magnetocrystalline anisotropy $E_A$ of a single domain particle is expressed as follows:

$$E_A = K\ D_S\ \sin^2\theta, \qquad (2)$$

where K is magnetocrystalline anisotropy constant, $D_S$ is size of nanoparticles, and θ is the angle between magnetization direction and easy direction of nanoparticles.[13] When thermal activation energy ($k_BT$, where $k_B$ is Boltzmann constant) is comparable with $E_A$, the direction of the magnetization of the nanoparticles starts fluctuate and goes through superparamagnetic relaxation. As the temperature increases over $T_B$, the magnetocrystalline anisotropy is overcome by the thermal energy, the direction of the magnetization of the nanoparticles follows the direction of the applied magnetic field which results in the superparamagnetic behavior of the nanoparticles. However, below $T_B$ thermal energy is no longer competent to overcome the magnetocrystalline anisotropy of the nanoparticles as a result the magnetization direction of the nanoparticles rotates from field direction to its own easy axis. In the present studied samples, the broad maxima is observed at a slightly lower temperature ($T_B$) than $T_{IRR}$ ($T_{IRR}$ is the temperature at which irreversibility start between ZFC and FC magnetization). This behavior results from a certain particle size distribution in system and reflects that a fraction of large particles freezes at $T_{IRR}$ whereas the majority of nanocrystallites are being blocked at $T_B$. The $T_B$ of Al doped $Ni_{0.2}Cd_{0.3}Fe_{2.5}O_4$ nanoparticles were found to increase from at 88 K to 110K with increase in Al doping. The increase in the $T_B$ reflects that size of the nanoparticles increases with Al doping which is in well agreement with XRD and TEM results. Moreover, it can be clearly seen from the magnetization versus field measurements that the magnetic moment was found to decrease with Al doping. The decrease in magnetic moment may be due to the dilution of the sub-lattice by the substitution of magnetic ions by the non-magnetic ions.[5] This decrease in magnetic moment infers that the Al ions occupies the octahedral site of the spinel lattice thereby decreasing the number of magnetic ions at B site and thus decreases the magnetic moment of the system that is evident from the magnetization measurements.

The near edge x-ray absorption fine structure (NEXAFS) study is most trustworthy method for the element specific characterization. The NEXAFS measurements were performed to investigate the electronic structure and chemical environment of Fe ions in $Ni_{0.2}Cd_{0.3}Fe_{2.5-x}Al_xO_4$ ($0 \leq x \leq 0.4$) ferrite nanoparticles. Fig. 4 shows the Fe $L_{3,2}$ –edge NEXAFS spectra of $Ni_{0.2}Cd_{0.3}Fe_{2.5-x}Al_xO_4$ ($0 \leq x \leq 0.4$) along with the reference compounds $Fe_2O_3$($Fe^{3+}$), FeO($Fe^{2+}$) and $Fe_3O_4$(Fe2+/Fe3+). The NEXAFS spectra at Fe $L_{3,2}$ –edge find out the 3d occupancy of the Fe ions and provide information about valence state of Fe ions in the studied samples. The Fe $L_{3,2}$ – edge spectra are primarily due to the Fe 2p-3d hybridization and are affected by the core-hole potentials. The intensity of these lines represents the total unoccupied Fe 3d states. The two broad multiple structures $L_3$ and $L_2$ observed in Fe $L_{3,2}$ spectra are well known for reference compounds $Fe_2O_3$, FeO, and $Fe_3O_4$. The main difference in the reference spectra can be clearly seen at the $L_3$ feature. The difference in $L_3$ features can be attributed to the variety of 3d electron configuration of Fe ions ($Fe^{2+}$ or $Fe^{3+}$) and the indication of local symmetry (tetrahedral or octahedral). The $L_3$ feature of $Fe_2O_3$ is characterized by a well developed doublet, a small intensity peak marked as A and a main peak marked as B, while in FeO the first peak becomes a shoulder of the main peak. These two spectral features in the $L_3$ region were assigned to Fe $t_{2g}$ and $e_g$ sub-bands, respectively. It can be clearly seen from the NEXAFS spectra that the spectral feature of $Ni_{0.2}Cd_{0.3}Fe_{2.5-x}Al_xO_4$ ($0 \leq x \leq 0.4$) nanoparticles resembles with $Fe_3O_4$ and FeO spectra indicates that Fe is in mix valence ($Fe^{3+}$ & $Fe^{2+}$) state. A difference spectra is also drawn (see inset in Fig. 4) by subtracting the x=0.0 spectra with respective spectra. It is observed that pre-edge peak is decreased with Al-ion doping and main peak is also shifted slightly, which shows an increase of $Fe^{2+}$ ions in the system. In order to obtain a better insight of the valence state of Fe ions, a linear combination fitting (LCF) at Fe K-edge NEXAFS spectra of the samples is carried out (see Fig. 5 (a) and (b)) from -20 to +40 eV taking as $Fe(Fe^0)$, $FeO(Fe^{2+})$, $Fe_2O_3(Fe^{3+})$ and $Fe_3O_4(Fe^{2+}/Fe^{3+})$ as standard spectra. The calculated parameters clearly show the existence of mixed valence state in $Ni_{0.2}Cd_{0.3}Fe_{2.5-x}Al_xO_4$ ($0 \leq x \leq 0.4$) nanoparticles.

## 4. CONCLUSIONS

We have successfully synthesis single phase polycrystalline nanoparticles of $Ni_{0.2}Cd_{0.3}Fe_{2.5-x}Al_xO_4$ ($0 \leq x \leq 0.4$) ferrites. XRD and TEM measurements indicate that particle size increases with Al doping. Magnetic measurements reflects that $Ni_{0.2}Cd_{0.3}Fe_{2.5-x}Al_xO_4$ ($0 \leq x \leq 0.4$) ferrites nanoparticles have the superparamagnetic behavior at room temperature and blocking temperature increases whereas magnetic moment decreases with Al doping. The NEXAFS measurements performed at Fe $L_{3,2}$ – edge and linear combination fitting (LCF) done at Fe K-edge spectra show that Fe is in mix valence state ($Fe^{3+}/Fe^{2+}$).

**Figure Captions**

Fig. 1 (Colour Online) X-ray diffraction patters of $Ni_{0.2}Cd_{0.3}Fe_{2.5-x}Al_xO_4$ ($0 \leq x \leq 0.4$) ferrite nanoparticles.

Fig. 2 (Colour Online) TEM micrograph of $Ni_{0.2}Cd_{0.3}Fe_{2.5-x}Al_xO_4$ ($0 \leq x \leq 0.4$) ferrite nanoparticles (a) x = 0.0, (b) x = 0.4. Upper left insets show the corresponding SAED pattern for x = 0.0 and 0.4. Upper right insets show the particle size distribution histograms for x = 0.0 and 0.4

Fig. 3 (Colour Online) Magnetization versus magnetic field measurements of $Ni_{0.2}Cd_{0.3}Fe_{2.5\_x}Al_xO_4$ ($0 \leq x \leq 0.4$) ferrite nanoparticles. Inset shows the ZFC and ZC curve for x = 0.0 and 0.4.

Fig. 4 (Colour Online) Normalized Fe $L_{3,2}$ -edge spectra $Ni_{0.2}Cd_{0.3}Fe_{2.5-x}Al_xO_4$ ($0 \leq x \leq 0.4$) ferrite nanoparticles plotted with reference spectra $FeO(Fe^{2+})$, $Fe_2O_3$ ($Fe^{3+}$) and $Fe_3O_4(Fe^{2+}/Fe^{3+})$ for comparison. Inset shows difference spectra obtained after subtracting x=0.0 spectra from respective spectra, to observe the difference with Al –ion doping.

Fig. 5 (Colour Online) (a) Normalized Fe K -edge spectra $Ni_{0.2}Cd_{0.3}Fe_{2.5-x}Al_xO_4$ ($0 \leq x \leq 0.4$) ferrite nanoparticles plotted with reference spectra $FeO(Fe^{2+})$, $Fe^2O^3$ ($Fe^{3+}$) and $Fe^3O^4(Fe^{2+}/Fe^{3+})$ for comparison. (b) Linear combination fitting (LCF) for x=0.0, 0.1, 0.2, 0.3 and 0.4 samples plotted with experimental spectra. Inset as table shows the calculated parameters.

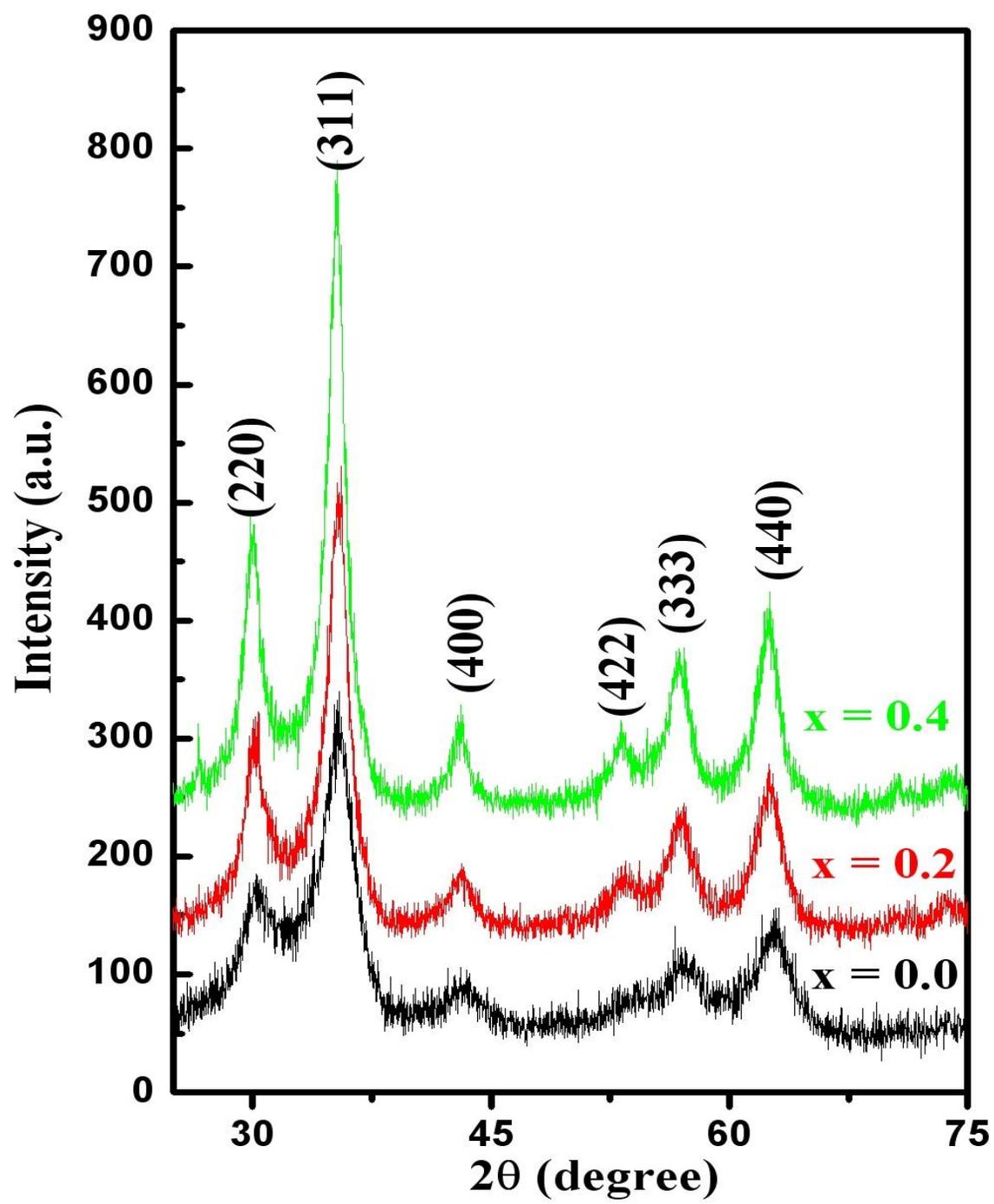

Fig. 1

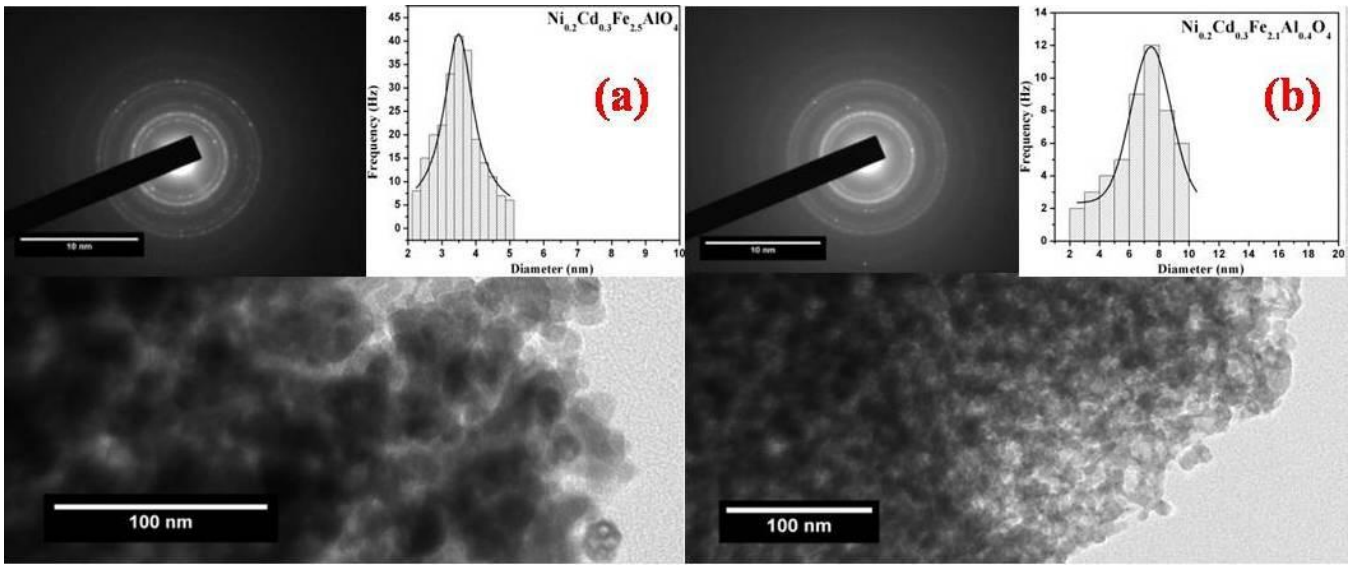

Fig. 2

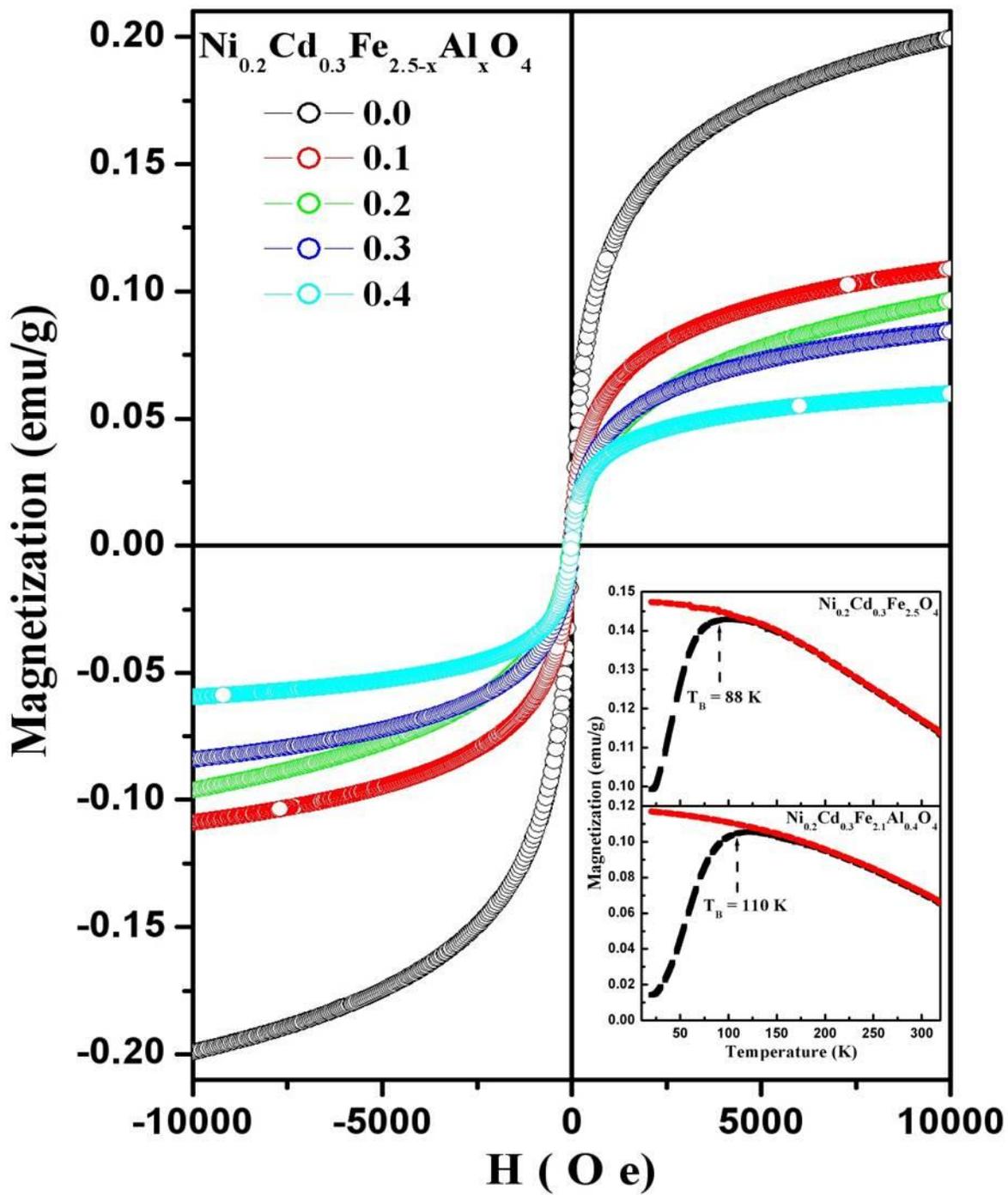

Fig. 3

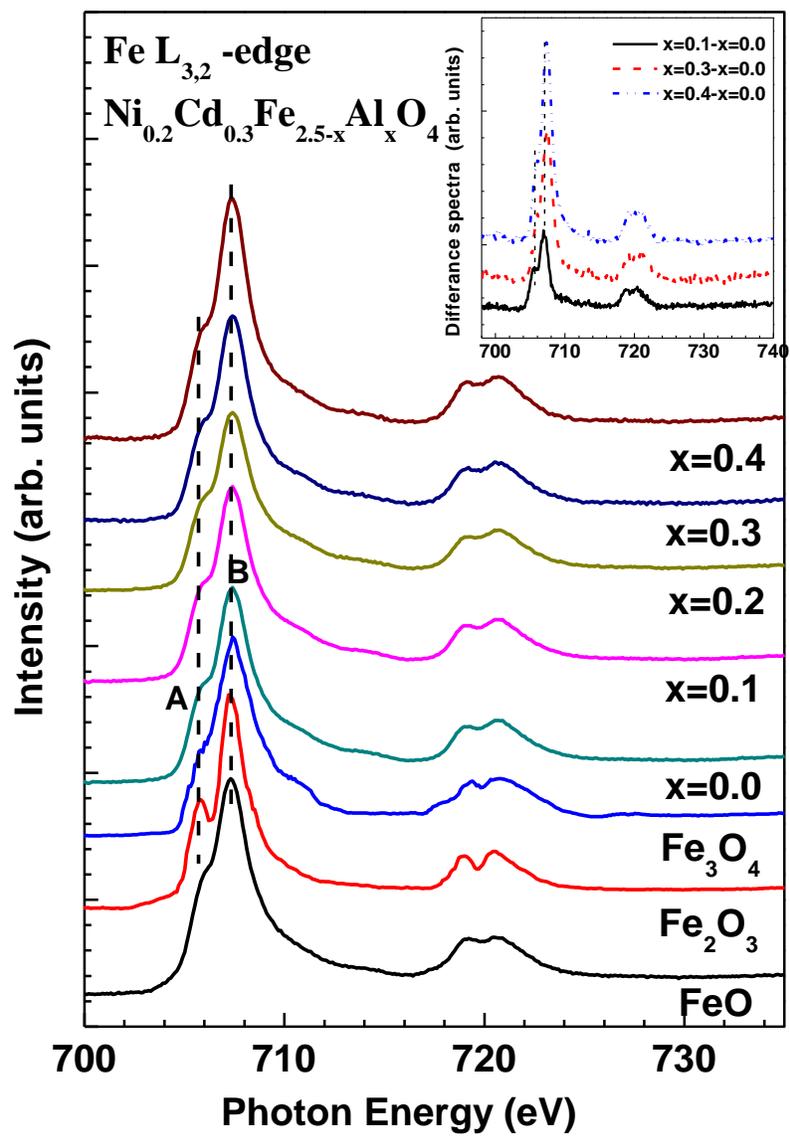

Fig. 4

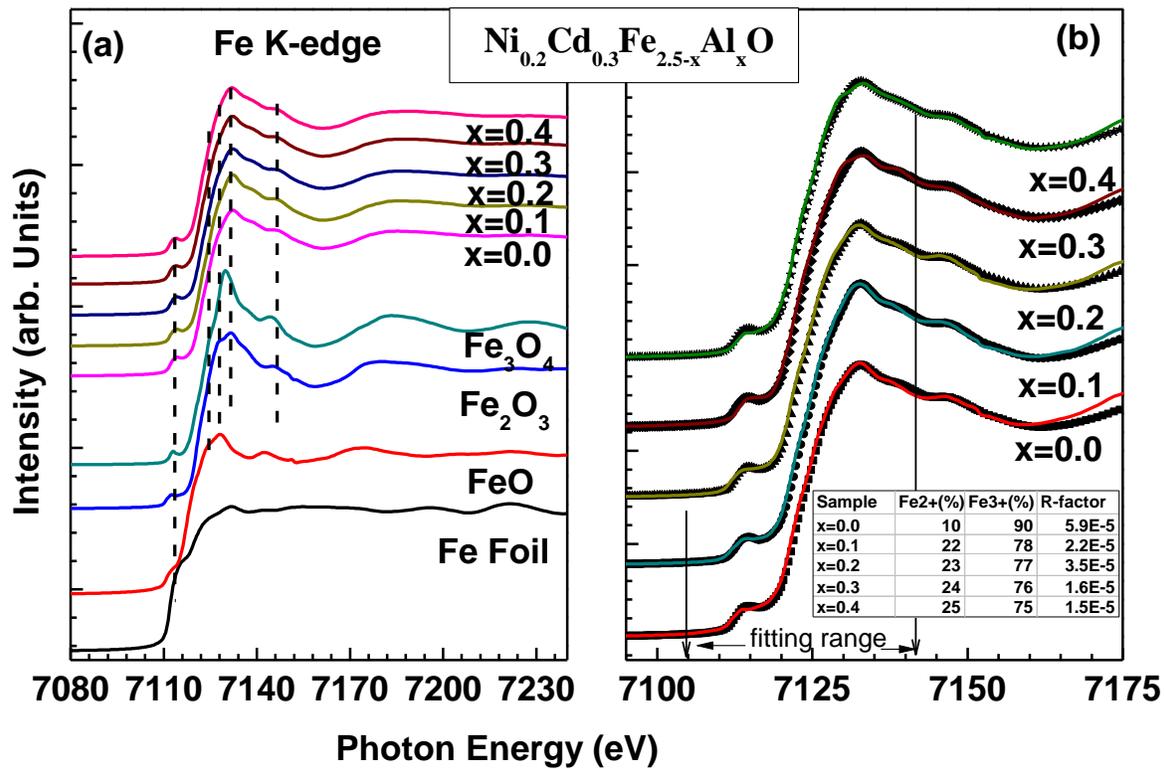

Fig. 5